\documentclass[twocolumn,prl]{revtex4-1}
\usepackage[latin9]{inputenc}
\usepackage{color}
\usepackage{amsmath}
\usepackage{amssymb}
\usepackage{graphicx}
\usepackage[unicode=true,pdfusetitle,
 bookmarks=true,bookmarksnumbered=false,bookmarksopen=false,
 breaklinks=false,pdfborder={0 0 0},pdfborderstyle={},backref=false,colorlinks=true]
 {hyperref}
\hypersetup{
 colorlinks,linkcolor=red,citecolor=blue}
\usepackage{breakurl}

\makeatletter

\usepackage{amsfonts}

\makeatother

\begin{document}

\title{Repetition rate tuning of soliton in microrod resonators}

\author{Rui Niu$^{1,3}$, Shuai Wan$^{1,3}$, Shu-Man Sun$^{1,3}$, Tai-Gao
Ma$^{1,3}$, Hao-Jing Chen$^{1,3}$, Wei-Qiang Wang$^{2,4}$, Zhizhou
Lu$^{2,4}$, Wen-Fu Zhang$^{2,4}$, Guang-Can Guo$^{1,3}$, Chang-Ling
Zou$^{1,3}$, Chun-Hua Dong$^{1,3,\dagger}$}

\affiliation{$^{1}$Key Laboratory of Quantum Information, CAS, University of
Science and Technology of China, Hefei, Anhui 230026, P. R. China}

\affiliation{$^{2}$State Key Laboratory of Transient Optics and Photonics, Xi\textquoteright an
Institute of Optics and Precision Mechanics (XIOPM), Chinese Academy
of Sciences (CAS), Xi\textquoteright an 710119, China}

\affiliation{$^{3}$CAS Center For Excellence in Quantum Information and Quantum
Physics, University of Science and Technology of China, Hefei, Anhui
230026, People's Republic of China}

\affiliation{$^{4}$University of Chinese Academy of Sciences, Beijing 100049,
China}

\date{\today}
\begin{abstract}
The coherent temporal soliton in optical microresonators has attracted
great attention recently. Here, we demonstrate the dissipative Kerr
soliton generation in a microrod resonator, by utilizing an auxiliary-laser-assisted
thermal response control. By external stress tuning, the repetition
rate of the soliton has been controlled over a large range of $30\,\mathrm{MHz}$.
Our platform promises precise tuning and locking of the repetition
frequency of coherent mode-locked comb in the microresonator, and
holds great potential for applications in spectroscopy and precision
measurements. 
\end{abstract}
\maketitle
Benefiting from the strongly enhanced nonlinear optics effects in
high-quality (Q) factor whispering gallery (WG) microresonators, frequency
combs have been observed in various experimental platforms via cascaded
nonlinear processes \cite{Kippenberg2011,DelHaye2007,Chembo2016,Guo2018,Pasquazi2018}.
In 2014, it was demonstrated that the coherent frequency comb, which
is called temporal soliton state, can be generated spontaneously by
the competition of Kerr nonlinearity, continuous laser driving and
dissipation in a crystalline WG microresonator \cite{Herr2013,Kippenberg2018}.
The dissipative Kerr soliton (DKS) offers broadband low-noise frequency
comb in frequency-domain, or the femotsecond pulse train in the time-domain.
Such DKS attracted great attention immediately \cite{Weiner2017},
and has been realized in various material and photonic platforms \cite{Wang2018,Gong2018,Yang2017,Cole2017},
and has also been extended to different frequency bands \cite{Brasch2016,Li2017,Lee2017,Pfeiffer2017}.
The DKS in microphotonic cavities has the merits of scalability, stability,
portability and low power consumption, thus it holds great potentials
for applications in ultrahigh data rate communication \cite{Marin-Palomo2017},
high precision optical ranging \cite{Trocha2018,Suh2018}, dual-comb
spectroscopy \cite{Suh2016}, low-noise microwave source \cite{Yi2015},
optical clock \cite{Papp2014} and astronomical spectrometer calibration
\cite{Obrzud2017}. 

Although great progress has been achieved on the proof-of-principle
demonstration of applications of soliton microcombs, there are still
many practical challenges when comparingthe soliton microcomb with
the commercial products based on bulky optics and fiber optics counterparts.
For instance, in the scenarios of time-keeping and precision spectroscopy
\cite{Cundiff2003}, the stabilization of the repetition rate ($f_{\mathrm{rep}}$)
of the soliton microcombs is essential. However, the $f_{\mathrm{rep}}$
varies from resonator to resonator, and is also sensitive to environment
temperature and pump power. Therefore, the efficient tuning of the
$f_{\mathrm{rep}}$ is on demand for soliton microcombs. In the past
few years, though the $f_{\mathrm{rep}}$ of incoherent frequency
comb in microcavities has been effectively tuned by various methods,
such as thermal tuning \cite{xue2016thermal}, mechanical tuning \cite{papp2013mechanical}
and pump frequency tuning \cite{huang2015low}, their applications
in soliton microcomb have not been demonstrated yet. 

In this letter, we experimentally demonstrate the effective tuning
of the repetition rate of DKS in a microrod cavity. By introducing
an auxiliary laser to excite the mode that is excluded from soliton
microcomb generations \cite{Geng2017,Geng:18,weiqiang2018}, the cavity
thermal response is effectively adjusted and the switching to the
DKS can be stably achieved. Based on such a controlling approach,
we realized the DKS in the microrod under different external bias
stress, and successfully tuned the repetition rate $f_{\mathrm{rep}}$
of the soliton microcomb over $30\,\mathrm{MHz}$. Our platform allows
the stabilization of the device temperature by the auxiliary laser,
precise and fast tuning of the $f_{\mathrm{rep}}$, thus making it
potential for locking soliton microcomb.

\begin{figure}
\includegraphics[width=1\columnwidth]{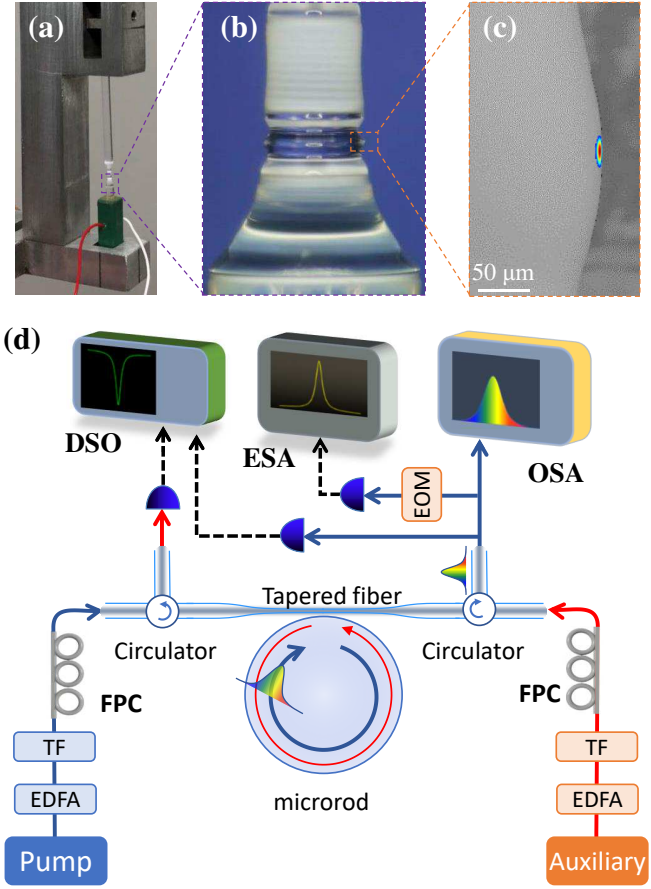} \caption{\label{fig:Fig1} (a) The setup for the mechanical tuning of microrod,
where a PZT (green block) is used to compress the microrod. (b) The
photo of the microrod cavity, with a diameter of about $1.2\,\mathrm{mm}$.
(c) The optical microscopy picture of microrod cavity and the corresponding
optical mode profile of the whispering gallery modes by numerical
simulation. (d) Schematic of the experimental setup of repetition
rate tuning. EDFA: erbium-doped fiber amplifier. TF: tunable filter.
FPC: fiber polarization controller. DSO: digital oscilloscope. OSA:
optical spectrum analyzer. ESA: electrical spectrum analyzer. EOM:
electro-optic modulator.}
\vspace{-6pt}
 
\end{figure}

Figure$\,$\ref{fig:Fig1}(a) shows the home made microrod resonator
and the associated mechanical tuning setup. The microrod cavity is
fabricated from a fused silica rod, where the surface is deformed
by heating the rod with a focused $\mathrm{CO_{2}}$ laser beam and
simultaneously rotating it \cite{papp2013mechanical,del2013laser}.
This microrod resonator for frequency comb applications was firstly
studied by P. Del\textquoteright Haye \cite{papp2013mechanical,del2013laser,Delbino:18},
promising an efficient approach to control the $f_{\mathrm{rep}}$.
By placing such a rod in our setup, its length and the stress in the
material can be changed by applying a voltage to the piezoelectric
transducer (PZT). An enlarged photo of the fabricated cavity is shown
in Fig.$\,$\ref{fig:Fig1}(b). There are two taper regions, while
the protrudent region in between is a WG cavity providing the confinement
potential in the vertical direction of the microrod. The electric
field intensity distribution of a typical fundamental WG modes is
shown in Fig.$\,$\ref{fig:Fig1}(c), with the cavity cross-section
profile taken from experimental data. The simulation result indicates
that light is confined at the equator, and the energy is very close
to the surface. When a voltage is applied to the PZT, the material
refractive index changes due to the stress and the cavity geometry
changes due to the strain, both lead to the changes of resonance frequency
and free spectral ranges (FSR). For a DKS microcomb, the repetition
rate directly depends on the FSR of the modes close to the pump laser
wavelength \cite{Herr2013}. Our experimental platform would efficiently
tune the FSR of the microrod, thus providing a fast and convenient
approach to control the repetition rate of a DKS microcomb.

\begin{figure}
\includegraphics[width=1\columnwidth,height=0.8\columnwidth]{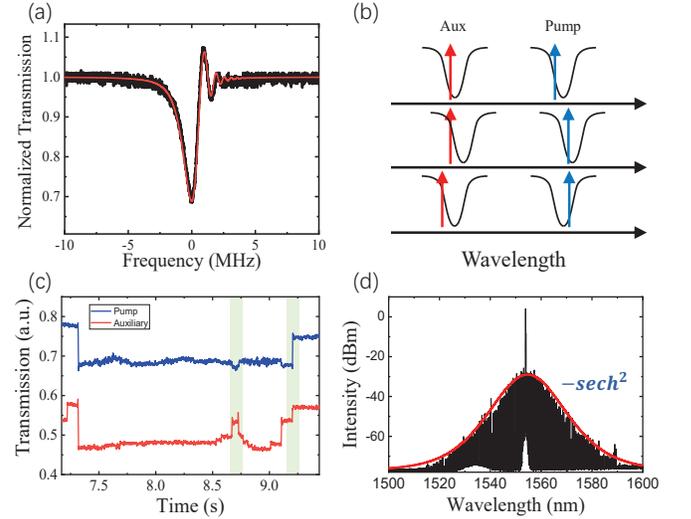}
\caption{(a) The typical optical mode spectrum at 1546nm which is measured
at sweeping speed $2.6\,\mathrm{MHz/\mu s}$. The red line represents
the theoretic fitting result with $Q_{0}=2\times10^{8}$. (b) Schematic
of auxiliary-laser-assisted thermal response control method. (c) Experimental
transmission of pump (blue) and auxiliary (red) lasers, the pump laser
settles on a low-noise soliton step. (d) Optical spectrum of single
DKS, the red line shows the spectral $\mathrm{sech^{2}}$ envelope.}
\label{fig:Fig2} 
\end{figure}

The experimental setup for soliton generation in the microrod resonator
is shown in Fig.$\,$\ref{fig:Fig1}(d), where a clockwise (CW) pump
laser (Toptica CTL 1550) is coupled into the microrod cavity through
a tapered fiber. The microrod used in our experiment has a diameter
of about $1.2\,\mathrm{mm}$, consistent with the observed FSR of
WG modes $\Delta_{\mathrm{FSR}}=55.6\,\mathrm{GHz}$, and the intrinsic
$Q$ factors of WG modes are about $2\times10^{8}$. A typical transmission
spectrum of the high-Q WGM by sweeping the laser wavelength is plotted
in Fig.$\,$\ref{fig:Fig2}(a), indicating a narrow linewidth of about
$\mathrm{MHz}$. Here, the laser power is very low to avoid the thermal
effect induced bistability. It is worth noting that the lineshape
of the resonance showing a ringing feature, which is a hallmark of
ultrahigh-Q resonance, as light could be stored inside the cavity
for a long duration of about $\tau=Q/\omega$ and interfere with the
transmitted sweeping laser \cite{Dong2009}.

In principle, our microrod resonator allows the formation of DKS as
long as the group velocity dispersion of the resonances is anomalous
around the working wavelength and the power of pump laser exceeds
the threshold \cite{Herr2013,Kippenberg2018}. Therefore, the microrod
resonator is fabricated to satisfy the condition of anomalous dispersion
at the telecom wavelength, and the input power of the  laser is amplified
by an erbium-doped fiber amplifier (EDFA, \textcolor{red}{upto 5$\,\mathrm{Watt}$}).
However, as demonstrated in other experimental platforms \cite{Guo2016,Yi2016},
in practical the thermal effect prevents us from directly generating
the DKS by simply sweeping the pump laser wavelength. To reconcile
this challenge, we adopt the thermal response control scheme by using
another auxiliary laser that couples the traveling wave WG resonator
in the counter-clockwise (CCW) direction \cite{Geng2017}, as depicted
by the red circuits in Fig.$\,$\ref{fig:Fig1}(d). The light in CW
(pump and generated comb) and CCW (auxiliary laser) directions are
separated by two circulators \cite{Geng:18,shen2018reconfigurable},
with the polarization of each laser is controlled by the fiber polarization
controllers (FPCs). And then the transmission data and optical spectrum
are recorded by the digital oscilloscope (DSO) and the optical spectrum
analyzer (OSA), respectively.

Figure$\,$\ref{fig:Fig2}(b) illustrates the mechanism of the auxiliary-laser-assisted
DKS generation \cite{Geng2017,Geng:18,weiqiang2018}. Firstly, the
pump and the auxiliary laser are settled at blue-detuning side of
the cavity mode. In the experiment, pump power and auxiliary power
were set at about $800\,\mathrm{mW}$. When the auxiliary laser drop
out off a cavity mode, the cavity cools down rapidly and the resonance
blue shifted, effectively scanning the pump to reach the DKS state.
Figure$\,$\ref{fig:Fig2}(c) shows the actual experimentally dynamics
of pump laser and auxiliary laser, the green region means that the
auxiliary laser launches a ``kick'' to the pump, only the latter
one triumphantly pushed the pump to the DKS state, while the success
rate mainly depends on the detuning of pump and auxiliary lasers.
By applying such a thermal response control method, we eventually
reached single soliton state, as shown in Fig. \ref{fig:Fig2}(d).

\begin{figure}
\includegraphics[width=1\columnwidth]{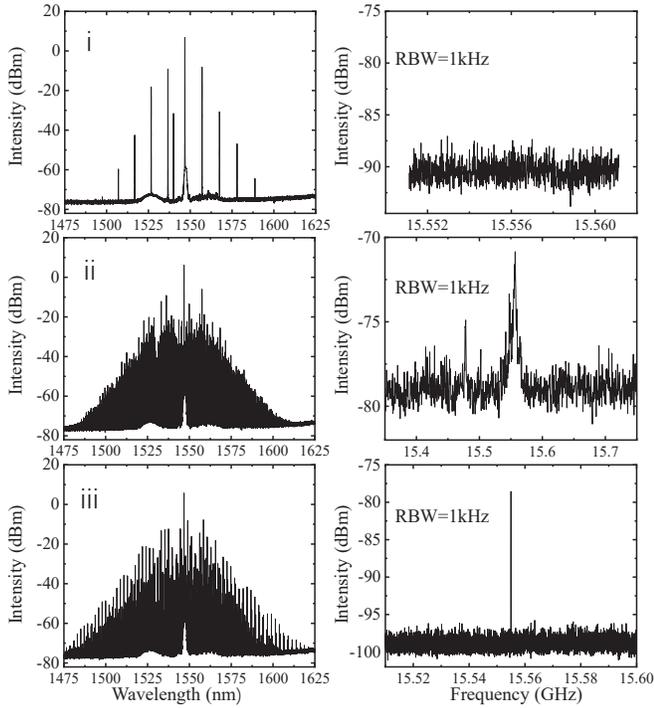}\caption{Optical spectra and RF beat note of primary comb state (i), MI state
(ii) and DKS state (iii). Here, the pump laser wavelength was set
at around 1546.8nm, and the auxiliary laser wavelength was set at
around 1540nm. the extra line near 1540nm in the primary comb spectrum
is the back scattered laser of the auxiliary laser. The resolution-bandwidth
(RBW) of the ESA is limited to 1kHz.}
\label{fig:Fig3} \vspace{-6pt}
\end{figure}

To validate the DKS microcomb generation in our experiments, we characterized
the output frequency combs in detail by resuming the optical spectrum
and the radio frequency (RF) beat note of comb lines. Limited by the
detection range of our detector and electrical spectrum analyzer (ESA),
we used an electro-optic modulator (EOM) modulated with frequency
of $\Omega=20\,\mathrm{GHz}$ to down convert the beat note signal
to less than $20\,\mathrm{GHz}$. Based on the preknowledge about
the range of FSR in our cavity, we can exactly determine the $f_{\mathrm{rep}}=f_{\mathrm{rep}}^{\prime}+2\Omega$
while $f_{\mathrm{rep}}^{\prime}$ is the measured frequency in ESA.
Typical results for different microcomb states achieved in our system
are plotted in Fig.$\,$\ref{fig:Fig3}. When the pump was blue detuned,
the microrod frequency comb exhibited low-noise primary comb state,
the corresponding RF signal of primary comb state was too large to
be detected due to the limited modulation depth of EOM (Fig.$\,$\ref{fig:Fig3}(a)).
When we precisely increased the pump laser wavelength, the high-noise
modulation instability (MI) state was achieved with single FSR spacing
while RF noise signal bandwidth was about $10\,\mathrm{MHz}$, as
shown in Fig.$\,$\ref{fig:Fig3}(b). Then, we carefully scanned the
auxiliary laser to precisely control the effective pump detuning,
thus realize the switching to soliton state with high probability.
For a typical multi-soliton state in Fig.$\,$\ref{fig:Fig3}(c),
the RF noise background is reduced by about $20\,\mathrm{dB}$ and
the noise peak bandwidth is also reduced by $20\,\mathrm{dB}$ (to
$100\,\mathrm{kHz}$) comparing to the MI state. 

\begin{figure}
\includegraphics[width=1\columnwidth]{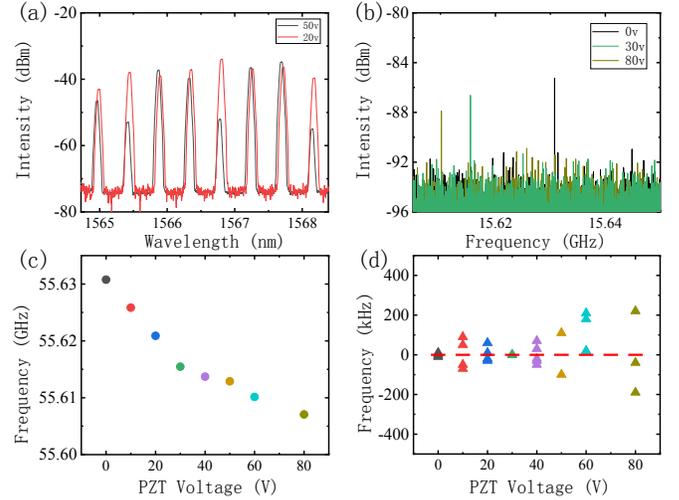} \caption{\label{fig:Fig4}(a) The optical spectra of soliton microcomb with
different PZT voltages. (b) The RF beat note with different PZT volatges.
(c) The dependence of RF beat note frequency with the different applied
PZT voltage. (d) The variation of the RF frequency for pump detunings
at the different applied PZT voltage.}
\vspace{-6pt}
 
\end{figure}

Since our system has the advantages of fast and convenient adjustment
of the structure, we demonstrate a mechanism for tuning the repetition
rate precisely through applying a mechanical force along the the vertical
direction of the microrod. In our experiment, we applied voltages
on the PZT, and adjusted the microrod cavity to critical coupling
at each voltage. Then we scanned the pump laser to 1557.2nm, by controlling
the auxiliary laser we got the same soliton state. Figure~\ref{fig:Fig4}(a)
shows the optical spectra of DKS microcomb under different voltage
on PZT. Due to the limited resolution of OSA ($0.02\,\mathrm{nm}$),
the offset of the comb lines for different voltages is hardly distinguished.
Therefore, the tuning of $f_{\mathrm{rep}}$ can only be accurately
inferred from RF spectra, as shown in Fig.$\,$\ref{fig:Fig4}(b).
Through increasing the applied voltage step by step, the repetition
rate variation was recorded and shown in Fig.$\,$\ref{fig:Fig4}(c),
which indicates the repetition rate reduces by nearly 30MHz with the
voltage applied to 80V, and the decreasing rate get slower after 40V.
The nonlinear slop of the $f_{\mathrm{rep}}$ tuning may attributed
to the nonlinear response of the PZT. The adjustable range of the
repetition rate is mainly restricted by the acceptable voltage of
PZT (120V at most) for the moment. It is also known that the repetition
rate of DKS also changes with the pump detuning, which might also
be used for $f_{\mathrm{rep}}$ tuning. To test this pump detuning-based
$f_{\mathrm{rep}}$ tuning method, we studied the $f_{\mathrm{rep}}$
at each voltage after reaching soliton state by slightly changing
the pump detuning. As shown in Fig.$\,$\ref{fig:Fig4}(d), at each
voltage the repetition rate varies $300\,\mathrm{kHz}$ at most, which
is much less than the variation caused by the PZT. As a consequence,
the repetition rate can be precisely changed over a range of $30\,\mathrm{MHz}$,
which is two orders larger than the pump-detuning approach. 

In conclusion, by introducing an auxiliary-laser-assisted thermal
response control, the generation of the DKS in an ultra high Q microrod
resonator is demonstrated successfully. And the tuning of the repetition
rate of the DKS in a microrod resonator is demonstrated by external
stress tuning. The repetition rate of the soliton state can be precisely
tuned over a broad band of $30\,\mathrm{MHz}$. Benefiting from the
advantage of precisely and fast tuning of repetition frequency of
DKS in the microresonator our platform can lock the repetition rate
stably, and provides a promising candidate for applications such as
metrology, spectroscopy and spectrometer calibration.
\begin{acknowledgments}
This work was supported by the National Key Research and Development
Program of China (Grant No.2016YFA0301303), National Natural Science
Foundation of China (Grant No.61575184, 11722436, 11874342 and 91536219),
and Anhui Initiative in Quantum Information Technologies (AHY130000),
the Fundamental Research Funds for the Central Universities. W.Q.W.,
Z.L. and W.F.Z. acknowledge the Strategic Priority Research Program
of the Chinese Academy of Sciences (Grant No. XDB24030600). This work
was partially carried out at the USTC Center for Micro and Nanoscale
Research and Fabrication.
\end{acknowledgments}

\end{document}